\documentclass[%
 reprint,
superscriptaddress,
showkeys,
 amsmath,amssymb,
 aps,
]{revtex4-2}

\usepackage{graphicx}
\usepackage{dcolumn}
\usepackage{bm}
\usepackage[hidelinks]{hyperref}

\usepackage{siunitx}  
\usepackage{comment}
\usepackage{float}
\usepackage{esvect}
\usepackage{gensymb}

\begin{document}

\preprint{APS/123-QED}

\title{Levitated macroscopic rotors with 10 hours of free spin at room temperature}

\author{Xianfeng Chen}
\email{xianfeng\_chen@imre.a-star.edu.sg}
\affiliation{A*STAR Quantum Innovation Centre(Q.InC), Agency for Science, Technology and Research(A*STAR), 2 Fusionopolis Way, 08-03 Innovis 138634, Singapore}
\affiliation{Institute for Materials Research and Engineering(IMRE), Agency for Science, Technology and Research(A*STAR), 2 Fusionopolis Way, 08-03 Innovis 138634, Singapore}

\author{Nirmala Raj}%
\affiliation{A*STAR Quantum Innovation Centre(Q.InC), Agency for Science, Technology and Research(A*STAR), 2 Fusionopolis Way, 08-03 Innovis 138634, Singapore}
\affiliation{Institute for Materials Research and Engineering(IMRE), Agency for Science, Technology and Research(A*STAR), 2 Fusionopolis Way, 08-03 Innovis 138634, Singapore}

\author{Ruvi Lecamwasam}
\affiliation{A*STAR Quantum Innovation Centre(Q.InC), Agency for Science, Technology and Research(A*STAR), 2 Fusionopolis Way, 08-03 Innovis 138634, Singapore}
\affiliation{Institute for Materials Research and Engineering(IMRE), Agency for Science, Technology and Research(A*STAR), 2 Fusionopolis Way, 08-03 Innovis 138634, Singapore}

\author{Mingxi Chen}
\affiliation{Institute for Materials Research and Engineering(IMRE), Agency for Science, Technology and Research(A*STAR), 2 Fusionopolis Way, 08-03 Innovis 138634, Singapore}

\author{Christina Yuan Ling Tan}
\affiliation{Institute for Materials Research and Engineering(IMRE), Agency for Science, Technology and Research(A*STAR), 2 Fusionopolis Way, 08-03 Innovis 138634, Singapore}

\author{Syed M. Assad}
\email{cqtsma@gmail.com}
\affiliation{A*STAR Quantum Innovation Centre(Q.InC), Agency for Science, Technology and Research(A*STAR), 2 Fusionopolis Way, 08-03 Innovis 138634, Singapore}
\affiliation{Institute for Materials Research and Engineering(IMRE), Agency for Science, Technology and Research(A*STAR), 2 Fusionopolis Way, 08-03 Innovis 138634, Singapore}

\author{Ping Koy Lam}
\email{pingkoy@imre.a-star.edu.sg}
\affiliation{A*STAR Quantum Innovation Centre(Q.InC), Agency for Science, Technology and Research(A*STAR), 2 Fusionopolis Way, 08-03 Innovis 138634, Singapore}
\affiliation{Institute for Materials Research and Engineering(IMRE), Agency for Science, Technology and Research(A*STAR), 2 Fusionopolis Way, 08-03 Innovis 138634, Singapore}
\affiliation{Centre for Quantum Computation and Communication Technology, Department of Quantum Science, Australian National University, Canberra, ACT 2601, Australia}

\date{\today}

\begin{abstract}
Low-dissipation rotors with large angular momentum are essential for precision sensing and probing macroscopic quantum phenomena. To date, low dissipation can only be achieved for micro-scale rotors. Here, we report a diamagnetically levitated millimeter-scale rotor exhibiting a measured dissipation rate as low as \SI{3.85}{\micro Hz} at room temperature, corresponding to a free spinning duration exceeding \SI{10}{hours}. The rotor is levitated stably over an axisymmetric permanent magnet trap, and can be driven up to \SI{930}{RPM} using contactless electrostatic actuation in high vacuum. 
Leveraging its low damping rate and large angular momentum, we realize a precision gyroscope with a measured sensitivity of \SI{6.5e-3}{\degree/s} and an estimated thermal-limited stability of \SI{5.7e-7}{\degree/\sqrt{h}}. These results establish diamagnetic levitation as a promising room-temperature platform for high-performance gyroscopes.      
\end{abstract}

\maketitle


\section{Introduction}
Levitation offers six degrees of freedom in unconstrained motion for rigid bodies, opening new frontiers in precision sensing and enabling new tests of macroscopic quantum physics~\cite{gonzalez2021levitodynamics}. For such applications, it is essential to reduce energy dissipation, which adds noise, reduces sensitivity, and decoheres quantum states. Recent advances in levitated optomechanics have demonstrated low dissipation in various levitation platforms~\cite{gonzalez2021levitodynamics,millen2020optomechanics}, including electrically levitated nanoparticles with dissipation rates as low as \SI{69}{nHz}~\cite{dania2024ultrahigh}, and optically levitated nanoparticles whose center-of-mass motion has been cooled down to the quantum ground state~\cite{delic2020cooling,piotrowski2023simultaneous}. Such systems exhibit extraordinary sensitivity in measuring forces~\cite{melo2024vacuum}, pressures~\cite{liu2024nanoscale}, accelerations~\cite{timberlake2019acceleration}, spin~\cite{jin2024quantum}, and magnetic field~\cite{ji2025levitated}.


The free rotation enabled by levitation has unique potential for probing macroscopic quantum phenomena~\cite{stickler2018probing,ma2020quantum,stickler2021quantum}, and detecting ultra-weak torques~\cite{ahn2020ultrasensitive}. Over the past decade, ultra-fast rotors with rotation speed from MHz to GHz have been realized~\cite{ahn2020ultrasensitive,ahn2018optically,kuhn2017optically,schuck2018ultrafast,monteiro2018optical}. 
Similarly, using electromagnetic forces, sub-millimeter steel spheres have been driven to extreme rotation speeds $>$ \SI{600}{kHz}~\cite{schuck2018ultrafast}. 

Common levitation methods such as optical tweezers or electromagnetic traps require a continuous power supply to sustain levitation, which induces extraneous noise and added system complexity. Moreover, while optical tweezers have made impressive strides towards mechanical quantum states, they are restricted to nano- or microscale particles. Access to the macroscopic regime is crucial for precision inertial sensors, where sensitivity scales with mass, as in gravimeters~\cite{rademacher2020quantum,abramovici1992ligo}, accelerometers~\cite{timberlake2019acceleration},
and gyroscopes~\cite{everitt2011gravity,poletkin2017mechanical}. Furthermore, low-dissipation macroscopic mechanical systems could offer a pathway toward exploring the quantum-classical boundary and testing models of quantum gravity~\cite{chen2013macroscopic,frowis2018macroscopic}.

An alternative means of levitation is provided by diamagnetic materials, which can stably levitate with zero energy consumption~\cite{simon2000diamagnetic}. One of the most strongly diamagnetic materials is pyrolytic graphite, allowing for levitodynamics of centimeter-scale systems suspended above permanent magnets at room temperature~\cite{chen2020rigid,chen2024nonlinear}. This offers a unique platform for sensitive macroscopic levitated sensors, and has been used from measurements of the Earth's tides~\cite{leng2024measurement} to exotic forces~\cite{yin2025experimental}. The key limitation of diamagnetic levitation is eddy currents, which are induced in graphite as it moves through the magnetic field, and cause strong dissipation. 
These can be suppressed by either engineering slits~\cite{romagnoli2023controlling,xie2023suppressing}, or using composites of graphite microparticles~\cite{chen2022diamagnetic,tian2024feedback}, enabling quality factor of $5\times 10^5$ in millimeter systems at room temperature. 

\begin{figure*}[!t]
\centering
\includegraphics[width=15cm]{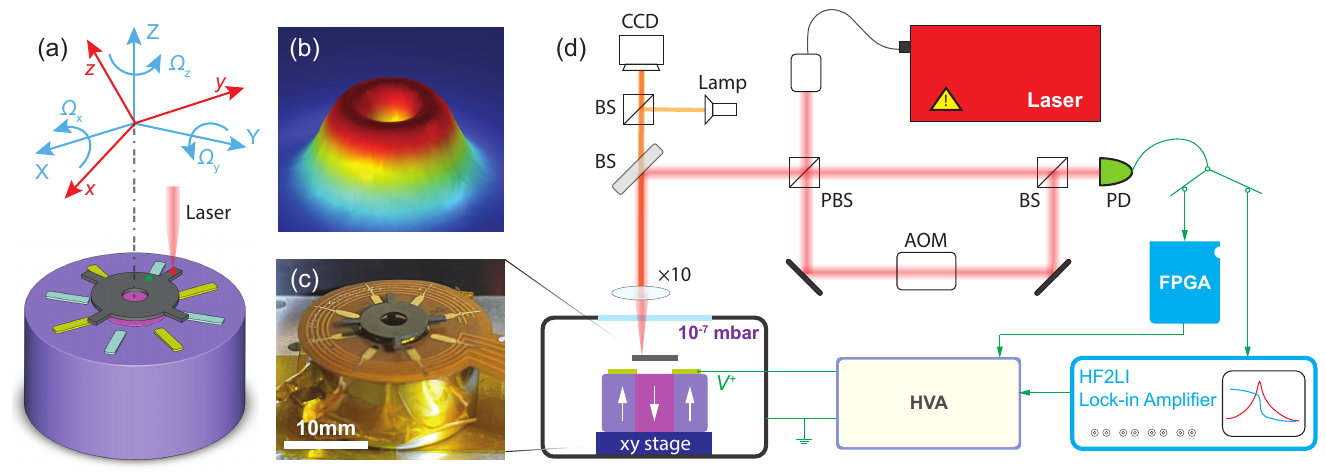}
\caption{\label{fig:setup} Experimental setup. (a) Schematic of a graphite rotor with four arms (gray) levitated above an inner cylindrical magnet (pink), and an outer ring magnet (purple). Eight electrodes are placed under the rotor, but only the gold ones are used to spin the rotor. The two permanent magnets have alternating magnetization. The laser is focused on the rotor arm (red dot) to trigger the driving voltage or measure the angular velocity; the laser is focused on the rotor's main body (green dot) when detecting its precession motion. Two coordinate systems are defined to represent the fixed frame (blue), and the rotating frame (red). (b) Simulated magnetic scalar potential at a plane \SI{1}{mm} above the magnets. Red color represents high potential, and blue indicates low. (c) A photograph of the levitation setup. Beneath the rotor is a customized PCB board with electrodes that generate electrostatic force to spin the rotor. Only four electrodes are used in this experiment. (d) Schematic of the optical readout system. The levitation setup is attached to an x-y piezo stage and placed inside a high vacuum chamber. To drive the rotor, signals generated from an FPGA or a lock-in amplifier are connected to the electrodes through a high-voltage amplifier (HVA) with 100 times amplification. To detect the rotor's motion, a laser Doppler vibrometer integrated with an imaging system is used to read out the rotor's velocity. BS: beam splitter; PBS: polarizing beam splitter; AOM: acousto-optic modulator; PD: photodetector; CCD: camera.}
\end{figure*}

The rotational dynamics of levitated graphite have been studied, predominantly for applications in bearingless motors and generators~\cite{khanna2010diamagnetically,su2015micromachined,xu2017realization,moser2001diamagnetic,ozawa2021principles}. To date, almost all experiments on rotational dynamics have been conducted in air, and with pyrolytic rather than composite graphite. Thus, low dissipation regimes relevant to precision measurements and quantum sensing schemes remain unexplored. Moreover, although diamagnetic levitation has been proposed as a promising approach for high-performance gyroscopes, there has not yet been systematic experimental validation.

Here, we levitate millimeter-scale diamagnetic rotors above a permanent magnet trap in an ultra-high vacuum environment. The axial symmetry of the magnetic potential permits unconstrained rotation of the rotor about the vertical axis. We use contactless electrostatic actuation to drive the rotors to high angular velocities and employ laser interferometry to measure their rotational dynamics precisely. By comparing the damping rates across different materials, dynamic modes, and pressures, we identify the dominant dissipation mechanisms and achieve a record-low dissipation rate for macroscopic rotors. Leveraging on both high angular momentum and low damping, we demonstrate a levitated gyroscope whose thermal-limited sensitivity is in the navigation grade.

\section{Results}

\subsection{Experimental setup}

\begin{figure}[h]
\includegraphics[width=8.6cm]{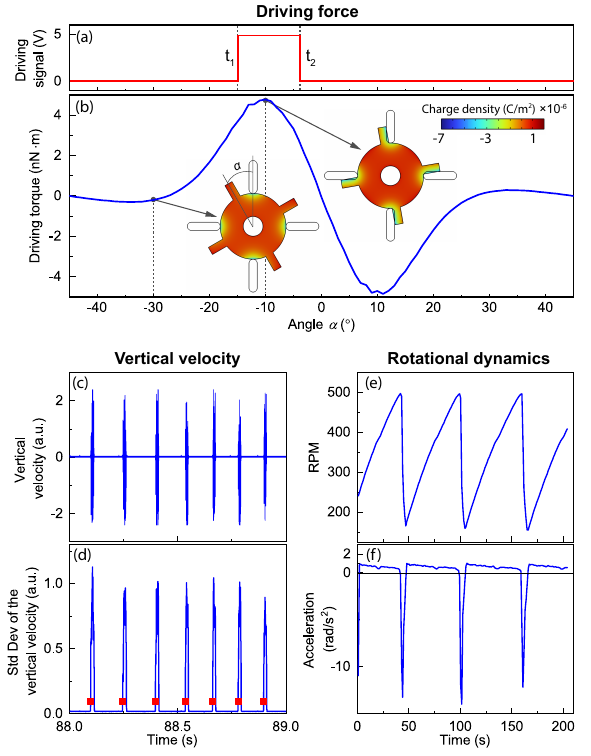}
\caption{\label{fig:RPM-get-torque} Driving force and rotational dynamics. (a) Driving signal as a function of the rotor position. (b) Simulated driving torque due to the electrostatic force with \SI{500}{V} as a function of the angle $\alpha$ between the rotor's arm and electrode. Angle $\alpha$ is defined as negative when the arm is on the left of the electrode. The torque is defined as positive when it is clockwise. Insets show the surface charge density distribution of the rotor at two angles, with the left-most rotor at $\alpha=-30\degree$, and the right-most at $\alpha=-10\degree$. (c) Vertical velocity detected by LDV. (d) Standard deviation obtained by processing the vertical velocity in Fig.~\ref{fig:RPM-get-torque}c using a rolling-window method. The dots are obtained by interpolating with a threshold, used to trigger the driving voltage (see Fig.~\ref{fig:RPM-get-torque}a) and compute the angular velocity. (e) Measured angular velocity of the rotor when continuously driven with the triggering signal. (f) Acceleration of the rotor.}
\end{figure}

Figure \ref{fig:setup} illustrates the levitation platform and its dynamic measurement system. The rotor, made from pyrolytic graphite, is laser-cut into a gear-shaped structure with four arms (see Fig.~\ref{fig:setup}a and \ref{fig:setup}c). Owing to the material's strong diamagnetism, the rotor levitates stably above two permanent magnets with a levitation gap approximately \SI{1}{mm}. This magnetic configuration constrains the rotor in five degrees of freedom, while permitting free rotation about the z-axis (Fig.~\ref{fig:setup}b). To drive the rotor, we place 4 planar electrodes beneath the rotor. The overlapping areas between the electrodes and the rotor arms generate a net driving torque on the rotor via electrostatic interaction. To drive the rotor efficiently, a feedback control strategy is implemented using an FPGA (see Fig.~\ref{fig:RPM-get-torque}a-b and Methods). Once rotation is initiated, we use a laser Doppler vibrometer (LDV) to monitor the rotor's dynamic response (Fig.~\ref{fig:setup}d). All measurements are conducted in high vacuum, at pressures below $\SI{1e-6}{mbar}$, to minimize air damping.

To determine the rotor's angular velocity, the LDV's laser spot is focused on one of the rotor's arms (see the red dot in Fig.~\ref{fig:setup}a). Each time the arms intersect the laser path, a distinct signal spike is produced (Fig.~\ref{fig:RPM-get-torque}c). Using the FPGA, we can transfer the velocity to standard deviation and extract the rotation speed by identifying the rising edges of the signal in real-time (Fig.~\ref{fig:RPM-get-torque}d). 

\subsection{Rotational dynamics}

For a rotor levitated in an axisymmetric magnetic-gravitational potential well, there is no restoring force for rotation around the $z$-axis. The rotational dynamics are governed by:
\begin{equation}
    I_\mathrm{z}\dot{\Omega}_\mathrm{z}+\mu\Omega_\mathrm{z}=\tau_\mathrm{e},
    \label{eq:rotational dynamics}
\end{equation}
where $I_\mathrm{z}$ is the moment of inertia of the rotor around $z$-axis, $\Omega_\mathrm{z}$ is the angular velocity, $\mu$ is the damping coefficient, and $\tau_\mathrm{e}$ is the electrostatic driving torque. According to Eq.~(\ref{eq:rotational dynamics}), the steady-state rotational velocity is determined by the balance between driving torque and damping. Once the driving force is removed, the motion decays exponentially:
\begin{equation}
    \Omega_\mathrm{z}=\Omega_0e^{-\gamma t},
    \label{eq:decay}
\end{equation}
where $\Omega_0$ is the initial angular velocity, and $\gamma = \mu/I_\mathrm{z}$ is the rotational damping rate, expressed in units of \SI{}{rad/s}.

Using the driving method and measurement setup (see Fig.~\ref{fig:setup} and Methods), we spin up the rotor in high vacuum to observe its dynamic response. In Fig.~\ref{fig:RPM-get-torque}e, we show a representative result for a rotor with a \SI{12}{mm} diameter, continuously driven with the signal in Fig.~\ref{fig:RPM-get-torque}a. The rotor accelerates gradually, reaching approximately \SI{500}{RPM}, before abruptly decelerating to below \SI{200}{RPM}. Under continuous driving, the dynamic cycle repeats periodically. 

Differentiating the angular velocity over time yields the rotor's angular acceleration, as shown in Fig.~\ref{fig:RPM-get-torque}f. During the spin-up phase, the rotor's acceleration decreases from approximately \SI{1.0}{rad/s^2} to \SI{0.5}{rad/s^2}, followed by a sharp transition to negative values. In high vacuum, the net torque on the rotor arises from two contributions: the electrostatic driving torque and the damping torque, which will be analyzed further in the following sections.

\subsection{Electrostatic driving torque}
The electrostatic force arises from capacitive interactions between the rotor's arms and the underlying electrodes. As this force is always attractive, a torque is generated when there is asymmetry in the overlapping area. In our levitation setup, high voltage signals ( typically above \SI{100}{V}) are applied to the electrodes (see Fig.~\ref{fig:setup}), creating a spatially varying electric field that induces a redistribution of charges on the conducting rotor. 

To analyze the driving torque quantitatively, we use finite element method (FEM) to simulate the electrostatic interaction between the rotor and electrodes. The electrostatic torque is computed as a function of the angle ($\alpha$) between a rotor arm and the electrodes when applied with +\SI{500}{V}, with the results shown in Fig.~\ref{fig:RPM-get-torque}b. The simulation reveals that the charges are redistributed on the rotor with negative charges in regions near the electrodes, while positive charges distribute over the remaining areas. The net charge on the rotor remains zero, as expected for an electrically floating conductor in vacuum. Moreover, the charge redistribution leads to different electrostatic torques at various angles. For instance, at $\alpha = -30 \degree$, the majority of negative charges reside on the rotor's main body, producing nearly zero net torque. In contrast at $\alpha = -10 \degree$, the negative charges accumulate near the rotor arms' edges, generating a significant torque. 

We can see that the electrostatic torque is periodic with rotor angle. The torque is only positive when $\alpha > -27 \degree$ and is maximum at $\alpha = -10 \degree$. Therefore, efficient rotor driving requires precise control of 
the driving signal to ensure that torque is consistently applied in the desired direction (see Methods). The peak simulated torque reaches several \si{nN \cdot m}, approximately one order of magnitude higher than the experimentally derived net torque \SI{0.464}{nN\cdot m} to \SI{0.23}{nN\cdot m} inferred from Fig.~\ref{fig:RPM-get-torque}f using Eq.~(\ref{eq:rotational dynamics}). This difference is expected given that $\tau_\mathrm{total}=\tau_\mathrm{e}-\mu \Omega_\mathrm{z}$, and that the net measured torque depends on both the effective actuation duration and the location of the laser spot used for triggering the actuation (see Methods). 

\subsection{Rigid body modes}
To investigate the origin of the sudden drop in rotational speed, we record the rotor’s behavior during spin-up using a high-speed camera. We find that when the rotor approaches \SI{500}{RPM}, the rotor starts to wobble dramatically and ultimately stops upon contacting the magnets. This critical speed \SI{500}{RPM} corresponds to a rotational frequency of $\SI{8.33}{Hz}$, suggesting a possible resonance with one of the rotor's rigid body modes.

\begin{figure}[h]
\includegraphics[width=8.6cm]{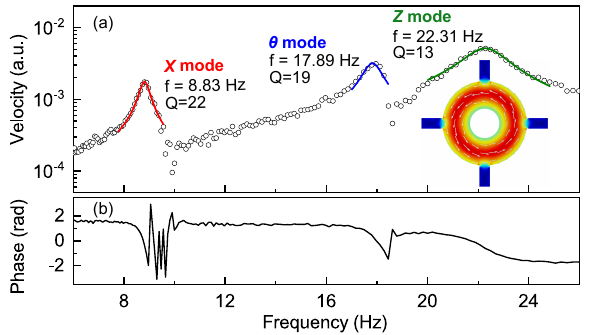}
\caption{\label{fig:frf-rigid} Frequency response curve of the levitated rotor. (a) Vibration amplitude as a function of driving frequency. The circle dots represent the measured data, while the solid lines represent the fitting with Lorentzian functions. Each peak's resonance frequency and Q factor are shown next to the peaks. The inset shows the simulated current density field (arrow surface) and the current density norm (red color represents high values). (b) Phase of the rotor's dynamic response.}
\end{figure}

To verify this, we characterize the rotor's rigid body dynamics by applying a combined DC and AC voltage to the electrodes while the rotor is levitated at rest (i.e., not spinning). Figure \ref{fig:frf-rigid} shows the detected amplitude and phase response of the rotor when sweeping the AC frequency from \SI{6}{Hz} to \SI{25}{Hz}. Three dominant resonance peaks are observed. Using a camera, we identify the three modes as: a horizontal mode (X mode), a librational mode ($\theta$ mode), and a vertical translational mode (Z mode). In principle, five rigid body modes are expected for a levitated plate~\cite{chen2020rigid}. The degenerate counterparts of the X and $\theta$ modes are not visible in Fig.~\ref{fig:frf-rigid}a, likely because the laser focus point used for detection has low sensitivity to those particular vibrations. By fitting the resonance peaks with Lorentzian functions, we extract their $Q$-factors, also shown in Fig.~\ref{fig:frf-rigid}a. In high vacuum, these $Q$s are dominated by eddy current damping~\cite{chen2020rigid,chen2022diamagnetic}. Interestingly, the $Q$s of the levitated rotor with a total diameter of \SI{12}{mm} (the outer diameter of the rotor main body is \SI{8}{mm}) are approximately 5 times lower than those measured for the levitated square plates~\cite{chen2020rigid}. This discrepancy can be attributed to differences in both geometry and magnetic configuration. In our case, the axisymmetric rotor–magnet arrangement enables freer eddy current circulation with almost all the currents flowing towards one direction, resulting in increased energy dissipation (see inset of Fig.~\ref{fig:frf-rigid}a).

\subsection{Energy dissipation}
We next investigate the energy dissipation associated with the levitated rotor's rotational mode. In diamagnetically levitated systems, damping in high vacuum is typically dominated by eddy currents~\cite{chen2020rigid}. Prior studies have shown that such losses can be significantly reduced by tailoring the geometry~\cite{romagnoli2023controlling,xie2023suppressing} or fabricating diamagnetic composites~\cite{chen2022diamagnetic,tian2024feedback,xiong2025achievement}. However, the damping mechanisms in their rotational modes remain poorly understood, despite prior demonstrations of diamagnetic levitation rotors~\cite{liu2008variable,su2015micromachined,xu2017realization}. 

\begin{figure*}[]
\includegraphics[width=14cm]{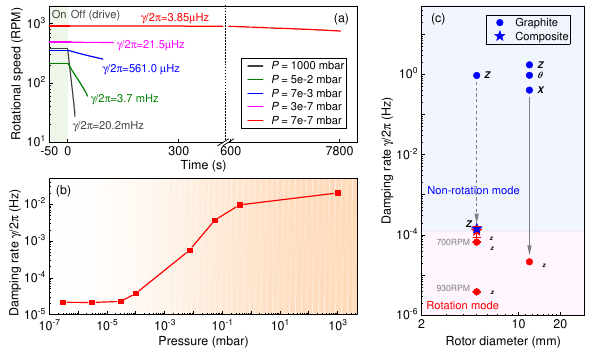}
\caption{\label{fig:ringdown-rotation}Energy dissipation of the levitated rotor. (a) Ringdown measurements of the rotor at different pressures and their damping rates obtained by fitting with exponential functions. The red line represents data from a \SI{5}{mm} rotor, while the others represent the data from a \SI{12}{mm} rotor. (b) Damping rate as a function of pressure for the \SI{12}{mm} rotor. (c) Comparison of the damping rates of rotors with different sizes, materials, and dynamic modes. Blue dots represent the non-rotation rigid body modes; red dots represent the rotational modes.}
\end{figure*}

To characterize rotational damping, we perform ringdown measurements in high vacuum. We first drive the rotor into a high rotation speed above \SI{200}{RPM}, then remove the drive and record the free decay of the angular velocity. Figure \ref{fig:ringdown-rotation}a shows representative decay traces at various pressures for a \SI{12}{mm} and a \SI{5}{mm} levitated rotor. The logarithmic decay curves exhibit linear behavior, indicating a damping torque proportional to $\Omega_\mathrm{z}$, consistent with the linear damping model $\mu \Omega_\mathrm{z}$. By fitting these decays with Eq.~(\ref{eq:decay}), we extract the damping rates $\gamma/2 \pi$, which are shown alongside the traces in Fig.~\ref{fig:ringdown-rotation}a. 

To investigate the pressure-dependent damping, we conduct measurements for the \SI{12}{mm} rotor over a wide pressure range from \SI{1000}{mbar} to \SI{1e-7}{mbar}. As shown in Fig.~\ref{fig:ringdown-rotation}b, the damping rate scales linearly with pressure in the intermediate regime $\SI{3e-1}{mbar}<P<\SI{3e-5}{mbar}$, confirming the dominance of residual gas damping. Below \SI{3e-5}{mbar}, the damping rate plateaus, falling three orders of magnitude below its atmospheric value, indicating a transition to pressure-independent, intrinsic dissipation. Notably, the damping rate of the rotational mode ($\Omega_\mathrm{z}$) is approximately five orders of magnitude lower than that of the non-rotational rigid body modes (X, $\theta$, and Z modes), as shown in Fig.~\ref{fig:ringdown-rotation}c.

In an ideal axisymmetric levitation system, rotational eddy-current damping should vanish. This is because, according to Faraday's law of induction, the electromotive force generated by a magnetic field is:
\begin{equation}
    \mathrm{emf}=\frac{\mathrm{d\Phi_{B}}}{\mathrm{d}t},
    \label{eq:faraday}
\end{equation}
where $\Phi_{B}$ is the magnetic flux through the rotor. For a perfectly balanced rotor spinning around its $z$-axis, the magnetic flux stays constant $\mathrm{d}\Phi_\mathrm{B}=0$, implying no eddy current generation. However, in practice, we observe residual damping even under high vacuum conditions. This is attributed to wobbling of the rotation axis, which breaks the symmetry and introduces time-varying magnetic flux $\mathrm{d}\Phi_\mathrm{B} \neq 0$, thereby generating eddy currents. 

To investigate this further, we fabricate two additional rotors and measure their damping rates in both the Z and $\Omega_\mathrm{z}$ modes. The first rotor is a smaller pyrolytic graphite rotor with a diameter $D=\SI{5}{mm}$. Due to a smaller size and moment of inertia, the rotor can be easily driven to a higher rotation speed. The rotor reaches a maximum speed of \SI{930}{RPM}, followed by a sudden drop due to resonance with a rigid body mode, similar to the larger rotor (Fig.~\ref{fig:RPM-get-torque}e). As shown in Fig.~\ref{fig:ringdown-rotation}c, ringdown measurements at initial speeds of \SI{700}{RPM} and \SI{930}{RPM} show that the damping rate at \SI{930}{RPM} is roughly one order smaller. At \SI{700}{RPM}, mild wobbling is observed, possibly due to coupling with a subharmonic of a rigid body mode or arising from the driving process. At \SI{930}{RPM}, the rotor spins more stably, resulting in reduced damping. 

The second rotor is fabricated by dispersing micro graphite particles in an epoxy matrix and cut into the same size as the small graphite rotor. As shown in Fig.~\ref{fig:ringdown-rotation}c, the Z-mode damping rate is reduced by approximately 4 orders of magnitude compared to the graphite rotor. This substantial reduction is because the epoxy isolates the graphite particles from each other, which significantly suppresses the eddy currents~\cite{chen2022diamagnetic}. However, the damping rate of the $\Omega_\mathrm{z}$ mode in the composite rotor does not decrease; instead, it increases slightly. This is due to greater wobble caused by lower intrinsic damping in the rigid modes. 

These results support the conclusion that rotational energy dissipation in diamagnetically levitated rotors under high vacuum primarily originates from eddy currents induced by wobbling motion. This wobble arises from imperfections in rotor geometry and dynamic coupling to rigid body modes (see also Fig.~\ref{fig:gyro}a). During spin-up, additional wobble can be introduced by asymmetric electrostatic actuation if the driving torque is not perfectly aligned with the rotation axis.

\section{levitated gyroscopes}
\begin{figure*}[]
\includegraphics[width=16cm]{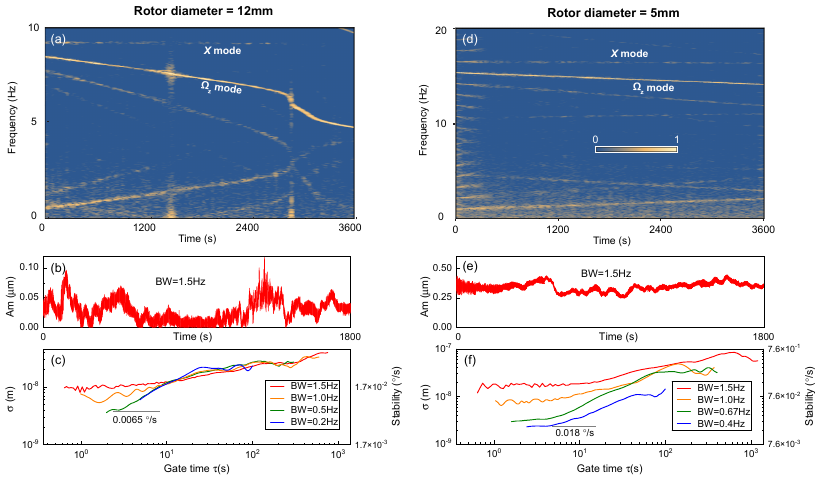}
\caption{\label{fig:gyro} Levitated gyroscopes. (a-c) Gyroscopic measurement results for the \SI{12}{mm} rotor. (a) Frequency spectrum of the rotor during its ringdown. The color represents the normalized amplitude on a logarithmic scale. (b) Amplitude fluctuation at the rotation frequency measured using the lock-in amplifier with a measurement bandwidth of BW=1.5Hz. (c) Allan deviation versus gate time of the amplitude measured with different bandwidths. (d-f) Gyroscopic measurement results for the \SI{5}{mm} rotor: (d) Spectrum versus time; (e) Amplitude versus time; (f) Allan deviation versus gate time.}
\end{figure*}
With their low damping and high moment of inertia, the levitated rotors serve as promising platforms for high-precision gyroscopic sensing. To evaluate their performance, we measure precession motion by focusing the laser on the edge of the rotor's main body (green dot in Fig.~\ref{fig:setup}a). The vertical displacement $Z$ encodes the precession signal, which depends on the input angular velocities $\Omega_\mathrm{X}$ and $\Omega_\mathrm{Y}$, as well as the rotor's RPM, size, and other physical parameters (see Supplementary Material for detailed modeling). All measurements are performed during the rotor's free ringdown to minimize influence from electrostatic drive. 

Figure \ref{fig:gyro} presents gyroscopic measurements for the two graphite rotors of diameters \SI{12}{mm} and \SI{5}{mm}. The larger rotor is driven up to \SI{510}{RPM} and the smaller rotor is driven up to \SI{930}{RPM}, before we switch off the electrostatic forces. For the larger rotor, Fig.~\ref{fig:gyro}a shows the frequency spectrum over time, obtained via a rolling-window FFT applied to a one-hour time-domain signal. A clear frequency trace, corresponding to the rotor's spin frequency, decaying from \SI{8.5}{Hz} to \SI{5}{Hz} is observed. During this process, a faint signature at around \SI{9}{Hz} shows up, which is the frequency trace of the X mode. The beating frequency traces $2\Omega_z-f_X$ (starting from \SI{7.87}{Hz}) and $f_X-\Omega_z$ (starting from \SI{0.73}{Hz}) can also be spotted. Notably, disturbances emerge around \SI{1500}{s} and \SI{2800}{s}. At \SI{2800}{s}, a sharp drop in spin frequency is observed, coinciding with the spin frequency approaching approximately one-third of the $\theta$ mode frequency and one-fourth of the $Z$ mode frequency (see also Fig.~\ref{fig:frf-rigid}a). This suggests a mode coupling and significant energy transfer to other modes~\cite{kecskekler2021tuning}.

In contrast, the smaller rotor exhibits a much cleaner spectrum (Fig.~\ref{fig:gyro}e), with the dominant frequency component closely tracking its rotational speed. The higher baseline RPM ($>\SI{14}{Hz}$) throughout the decay process keeps it away from low-frequency noise and modal coupling. Some beating traces are observed in the first 5 minutes, attributed to driving torque, but they dissipate quickly, suggesting effective self-stabilization of the rotor.

To assess the stability of the levitated rotors, we use a lock-in amplifier to track the precession amplitude at their rotational frequencies. Since the rotation speed decreases with a small decaying rate, the measurement bandwidth and duration are tuned carefully to cover the frequency span. Time traces of the precession amplitude are shown for both large and small rotors in Fig.~\ref{fig:gyro}b and Fig.~\ref{fig:gyro}e, respectively. The corresponding Allan deviation curves with different measurement bandwidths are shown in Fig.~\ref{fig:gyro}c and Fig.~\ref{fig:gyro}f. Unlike conventional Allan deviation plots, which decrease at short gate times and increase at longer ones~\cite{chen2021diamagnetically,xiong2025achievement}, we observe a monotonic increase, indicating that long-term drift dominates the stability. The primary sources of drift include temperature fluctuation, low-frequency mechanical vibrations, and the open-loop decay of spin frequency. As expected, Allan deviation improves with narrower measurement bandwidth due to noise suppression. 

From the gyroscope modeling (see Supplementary Materials), we derive the system scale factor relating precession amplitude to angular input of \SI{3.34e-5}{m/(rad/s)} for the large rotor at \SI{510}{RPM} and \SI{7.53e-6}{m/(rad/s)} for the small rotor at \SI{930}{RPM}. Using these values, we convert the Allan deviation to units of \si{\degree/s}, plotted on the secondary axis of Fig.~\ref{fig:gyro}c and f. The lowest measured angular resolution is \SI{0.0065}{\degree/s} for the large rotor, approximately one order of magnitude better than the small rotor. Although the small rotor achieves slightly lower displacement-based Allan deviation, its scale factor is significantly smaller, resulting in reduced angular sensitivity. 

\section{Discussion and conclusions}
Levitated rotors with different levitation techniques have been demonstrated at fast rotation speeds and proposed for developing gyroscopes~\cite{kuhn2017optically,ahn2018optically,zielinska2024long,ahn2020ultrasensitive,schuck2018ultrafast}. In this work, we demonstrate, for the first time, a diamagnetically levitated millimeter-scale gyroscope operating at room temperature. The performance of the \SI{12}{mm} graphite rotor, spinning at \SI{510}{RPM}, surpasses that of a smaller \SI{5}{mm} rotor at \SI{930}{RPM}, highlighting the crucial role of rotor size in enhancing inertial sensor performance, i.e., larger moment of inertia yields higher angular momentum and thus improved sensitivity.

\begin{figure}[]
\includegraphics[width=8cm]{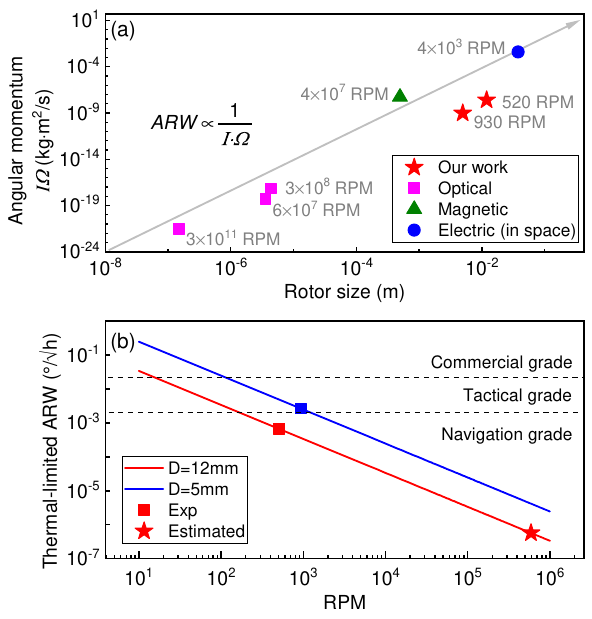}
\caption{\label{fig:gyro-benchmark}(a) Angular momentum versus size of different levitation rotors (optical~\cite{ahn2020ultrasensitive,zeng2024optically,arita2013laser}, magnetic~\cite{schuck2018ultrafast} and electric~\cite{everitt2011gravity}). Note that the electrically levitated rotor can only be implemented in space. (b) Thermal-limited angular random walk ($ARW$) as a function of RPM estimated using the system parameters and the achieved RPM (square dots) as well as the predicted ultimate RPM (circle dot).}
\end{figure}

The best achieved angular resolution, 
\SI{0.0065}{\degree/s} in our experiments, is one order of magnitude lower than the first optically levitated gyroscope operating at MHz speeds (\SI{0.08}{\degree/s})~\cite{xiong2025achievement}. This enhancement is due to the rotor's angular momentum, as gyroscope sensitivity is fundamentally limited by thermal noise and scales inversly with $I \Omega$~\cite{poletkin2017mechanical} (see also Eq.~~(\ref{eq:arw})). As summarized in Fig.~\ref{fig:gyro-benchmark}a, although optically levitated rotors achieve ultrahigh rotation speeds, their nanoscale dimensions limit their angular momentum. In contrast, the electrostatically levitated rotor from the Gravity Probe B project achieved the highest known angular momentum and sensitivity~\cite{everitt2011gravity}, but requires operation in space. Among ground-based methods, magnetic levitation has a large angular momentum. Particularly, diamagnetic levitation offers a unique advantage: zero power consumption, which minimizes noise from active stabilization and allows passive, long-term operation.

The gyroscope demonstrated here reaches commercial-grade performance~\cite{passaro2017gyroscope}, and it shows strong potential for further advancement. The ultimate limit is set by thermal noise, expressed as the angular random walk ($ARW$)~\cite{poletkin2017mechanical}:
\begin{equation}
    ARW=\frac{\sqrt{4 k_\mathrm{B} T \mu_\mathrm{\theta}}}{I_\mathrm{z} \Omega_\mathrm{z}}\frac{180\cdot 60}{\pi} \quad (\degree /\sqrt{h}), 
    \label{eq:arw}
\end{equation}
where $k_\mathrm{B}$ and $T$ are the Boltzmann constant and temperature, respectively; $\mu_\mathrm{\theta}$ is the damping coefficient of the rotor's $\theta$ mode. Using measured parameters, we estimate the thermal-limited $ARW$ (Fig.~\ref{fig:gyro-benchmark}b) and show that the large rotor already operates in the navigation-grade range~\cite{el2020inertial}. Based on the estimated driving torque (Fig.~\ref{fig:RPM-get-torque}), the rotor is theoretically capable of reaching rotation speeds approaching \SI{1e6}{RPM}. Such high speeds can be realized by feedback cooling the rigid body modes, as demonstrated in~\cite{schuck2018ultrafast}, and would reduce low-frequency vibrational noise significantly. Further improvements in rotor mass balance and closed-loop control~\cite{IEEE-STD-813-1988} could suppress wobble and long-term drift, pushing performance toward the thermal-noise limit of \SI{5.7e-7}{\degree/\sqrt{h}}.

Beyond gyroscopic applications, the levitated rotor exhibits an exceptionally low energy dissipation rate of \SI{3.85}{\micro Hz} at room temperature—the lowest ever recorded for a mechanical rotor at millimeter scale, and comparable to the best-performing nanoscale rotors (\SI{1.6}{\micro Hz})~\cite{monteiro2018optical}. This enables continuous, undriven rotation for over 10 hours. We show that the dominant source of dissipation is eddy current damping induced by rotational wobble, as confirmed by comparing ringdown behavior of graphite and composite rotors. While composite materials reduce damping in translational modes by limiting eddy current loops~\cite{chen2022diamagnetic}, the damping in rotation persists due to wobble-related flux changes.

In summary, we investigate the rotational dynamics and energy dissipation of millimeter-scale rotors levitated via diamagnetic force in high vacuum. Using electrostatic actuation and FPGA-based control, we achieve rotation speeds up to \SI{930}{RPM}, limited by the excitation of rigid body resonances. Ringdown measurements reveal that rotational damping is nearly five orders of magnitude lower than that of non-rotational modes, with a minimum dissipation rate of \SI{3.85}{\micro Hz}. The remaining dissipation is due to eddy currents induced by mechanical wobble, which can be minimized through precision rotor balancing. Finally, we demonstrate a diamagnetic gyroscope with a measured sensitivity of \SI{6.5e-3}{\degree/s}, and show that its thermal-limited $ARW$ may reach \SI{5.7e-7}{\degree/\sqrt{h}}, marking a significant step toward benchmark, high-performance inertial sensors and quantum optomechanical platforms.

This study establishes that diamagnetically levitated rotors can achieve exceptionally low dissipation at macroscopic scales—beyond the reach of existing optical levitation—paving the way for high-performance gyroscopes~\cite{zeng2024optically}, ultrasensitive torque sensors~\cite{ahn2020ultrasensitive}, and experimental platforms for testing macroscopic quantum phenomena~~\cite{stickler2021quantum, ma2020quantum, gonzalez2021levitodynamics,chen2013macroscopic,frowis2018macroscopic,aspelmeyer2014cavity, stickler2021quantum, kuhn2017full}.


\section{Methods}
\textit{Sample preparation:} Two diamagnetic materials were employed in this study. The first is a pyrolytic graphite sheet, mechanically polished to the desired thickness to minimize surface roughness. The second is a diamagnetic composite, fabricated following the procedure described in our previous work~\cite{chen2022diamagnetic}, by dispersing micro-scale graphite particles (Nanografi Nano Technology) into an epoxy matrix. The graphite particles have a nominal size of 1–\SI{5}{\micro\meter}, and the mass ratio of graphite to epoxy is $35\%$. Both materials were patterned into four-armed rotor geometries using a precision micro laser cutter, as shown in Fig.~\ref{fig:setup}c.

\textit{Driving strategy:}
As illustrated in Fig.~\ref{fig:setup}a, the levitated rotors are actuated via electrostatic forces using four electrodes placed beneath them. These electrodes are patterned on custom-designed printed circuit boards (JLCPCB) and are powered through a $\times$100 high-voltage amplifier (Trek Model 2210).

To initiate rotation, we manually toggle the electrodes until the rotor begins spinning. Once rotation is established, an automated control program developed in LabVIEW and deployed on a National Instruments FPGA is activated.

As shown in Fig.~\ref{fig:setup}a and Figs.~\ref{fig:RPM-get-torque}a–d, when a rotor arm intersects the laser beam, a voltage spike is detected by the system. This event triggers the FPGA to activate the electrodes for a short duration ($t_1$ to $t_2$ in Fig.~\ref{fig:RPM-get-torque}a), corresponding to the angular span during which the arm is directly under the laser—approximately \SI{11.5}{\degree} for the large rotor and \SI{13.6}{\degree} for the small rotor.

The choice of trigger angle $\alpha_0$ is crucial for efficient rotation, as demonstrated in Figs.~\ref{fig:RPM-get-torque}a-b. For instance, triggering at $\alpha_0 = \SI{-40}{\degree}$ results in nearly zero net torque, while triggering at $\alpha_0 = \SI{-6}{\degree}$ yields almost zero net work over the actuation period. Effective synchronization between detection and actuation ensures optimal electrostatic torque generation and sustained rotor acceleration.

\vspace{1cm}
\section*{Acknowledgments}
The authors thank Kin Seng Lai and Wei Hsiung Wee from DSO National Laboratories for valuable discussions. We also gratefully acknowledge Youngwook Cho, Tao Wang, Chenyue Gu, Biveen Shajilal, Mengting Jiang, and Xuezhi Ma for their assistance in constructing the experimental setup. This research/project is supported by the DSO National Laboratories (No: DSOCL23081) and the National Research Foundation, Singapore, under its Competitive Research Programme (CRP Award No: NRF-CRP30-2023-0002). Any opinions, findings and conclusions or recommendations expressed in this material are those of the author(s) and do not reflect the views of National Research Foundation, Singapore.

\vspace{1cm}
\section*{Author contributions}
X.C. and P.K.L. conceived the project. X.C., N.R. and S.M.A performed the experiments. N.R. and S.M.A designed and implemented the control program. X.C. and R.L. developed the theoretical modeling. M.C. and C.Y.L.T. fabricated the composites and carried out the laser cutting. X.C., N.R., S.M.A and P.K.L analyzed and interpreted the results. X.C., S.M.A and P.K.L secured the funding and supervised the project. X.C. wrote the manuscript with input from all authors. All authors have read and approved the final manuscript.

\bibliography{rotor}

\section*{Supplementary information}
Here we describe the modeling of a levitated 2-axis rate gyroscope, adapted from reference~\cite{poletkin2017mechanical}.
The coordinate systems are defined as shown in Fig.~\ref{fig:setup}a. XYZ is the fixed coordinate frame. xyz is the coordinate frame of the rotor. $x_r y_r z_r$ is the rotating frame with a speed of $\Omega_Z$ with respect to XYZ, which are not shown in the figure.

The equations of motion expressed by the two generalized rotational coordinates $q_x, q_y$ for the gyroscope are given by:

\begin{equation}
    \begin{cases}
        I_x \ddot{q}_x + \mu_x \dot{q}_x + C_{dx} q_x - h_y \dot{q}_y = \tau_x; \\
        h_x \dot{q}_x + I_y \ddot{q}_y + \mu_y \dot{q}_y + C_{dy} q_y = \tau_y,
    \end{cases}
    \label{eq: eom_rotation}
\end{equation}
where the torques acting on the $x$ and $y$ axes of the rotor are given by:

\begin{equation}
    \begin{cases}
        \tau_x = \tau'_Y \sin \Omega_Z t - \tau'_X \cos \Omega_Z t; \\
        \tau_y = \tau''_Y \cos \Omega_Z t + \tau''_X \sin \Omega_Z t,
    \end{cases}
\end{equation}
where

\begin{equation}
    \begin{aligned}
        \tau'_X &= (I_z + I_x - I_y) \Omega_Z \cdot \Omega_Y + I_x \dot{\Omega}_X; \\
        \tau'_Y &= (I_z + I_x - I_y) \Omega_Z \cdot \Omega_X - I_y \dot{\Omega}_Y; \\
        \tau''_X &= (I_z - I_x + I_y) \Omega_Z \cdot \Omega_Y + I_x \dot{\Omega}_X; \\
        \tau''_Y &= (I_z - I_x + I_y) \Omega_Z \cdot \Omega_X - I_y \dot{\Omega}_Y.
    \end{aligned}
\end{equation}

$\Omega_X$ and $\Omega_Y$ are the input angular rates. The parameters $h_x$ and $h_y$ are given by:

\begin{equation}
    h_x = I_x (1 - \kappa_x) \Omega_Z, \\ h_y = I_y (1 - \kappa_y) \Omega_Z,
\end{equation}
where

\begin{equation}
    \kappa_x = \frac{I_z - I_y}{I_x}, \quad \kappa_y = \frac{I_z - I_x}{I_y}.
\end{equation}
These are the respective gyroscope constructive parameters relative to the $x$ and $y$ axes. The damping terms $C_{dx}$ and $C_{dy}$ include contributions from dynamic stiffness:
\begin{equation}
    C_{dx} = c_s + \kappa_x I_x \Omega_Z^2, \quad C_{dy} = c_s + \kappa_y I_y \Omega_Z^2,
\end{equation}
where $c_s$ represents the angular position stiffness produced by a contactless suspension holding and rotating the rotor.

The behaviour of the gyroscope with respect to the fixed coordinate frame (CF) is described by the angular generalized coordinates, $\beta_X$ and $\beta_Y$. These coordinates have the following relationship to $q_x$ and $q_y$:

\begin{equation}
    \begin{cases}
        \beta_X = q_x \cos\Omega_Z t - q_y \sin\Omega_Z t; \\
        \beta_Y = q_x \sin\Omega_Z t + q_y \cos\Omega_Z t.
    \end{cases}
    \label{eq:beta_relations}
\end{equation}

In our experiments, we can only measure the $Z$-direction displacement of a point on the rotor. Now, we derive the position of a point on the edge of a spinning rotor.

A point at the edge of the rotor, which has a radius $R$, is described in the rotor's local coordinate system $(x, y, z)$ as:

\begin{equation}
    P_r = \begin{bmatrix} R \cos(\Omega_Z t) \\ R \sin(\Omega_Z t) \\ 0 \end{bmatrix}.
\end{equation}

To express this point in the fixed coordinate frame $XYZ$, we apply rotations about the $X$ and $Y$ axes. First, a rotation by $\beta_X$ around the $X$-axis:
\begin{equation}
    R_X(\beta_X) = \begin{bmatrix} 1 & 0 & 0 \\ 0 & \cos\beta_X & -\sin\beta_X \\ 0 & \sin\beta_X & \cos\beta_X \end{bmatrix}
\end{equation}
Then, a rotation by $\beta_Y$ around the $Y$-axis:
\begin{equation}
    R_Y(\beta_Y) = \begin{bmatrix} \cos\beta_Y & 0 & \sin\beta_Y \\ 0 & 1 & 0 \\ -\sin\beta_Y & 0 & \cos\beta_Y \end{bmatrix}
\end{equation}
The total transformation matrix is:
\begin{equation}
    R = R_Y(\beta_Y) R_X(\beta_X)
\end{equation}

Applying the transformation:

\begin{equation}
    P_X = R P_r,
\end{equation}
and expanding this multiplication, the final coordinates of the point in the fixed frame are obtained:

\begin{align}
    \begin{cases}
    X &= R \cos\beta_Y \cos(\Omega_Z t) + R \sin\beta_Y \sin\beta_X \sin(\Omega_Z t) \\
    Y &= R \cos\beta_X \sin(\Omega_Z t) \\
    Z &= - R \sin\beta_Y \cos(\Omega_Z t) + R \cos\beta_Y \sin\beta_X \sin(\Omega_Z t)        
    \end{cases}    
    \label{Eq: Z}
\end{align}

The displacement $Z$ is derived by substituting Eq.~ (\ref{eq: eom_rotation}) and Eq.~ (\ref{eq:beta_relations}) into Eq.~ (\ref{Eq: Z}). Using the measured parameters for the rotors in our experiments, we can obtain the scale factors relating the displacement $Z$ to the input angular rates $\Omega_X$ and $\Omega_Y$.

\end{document}